\documentclass
[superscriptaddress,secnumarabic,amssymb,amsmath,nobibnotes,aps,prd,showkeys,showpacs,nofootinbib,onecolumn,notitlepage]{revtex4-1}
\usepackage{amsmath}
\usepackage{amssymb}
\usepackage{tensor}
\usepackage{latexsym}
\usepackage{enumerate}
\usepackage{bbm}
\usepackage{amsthm}
\usepackage{graphicx}
\usepackage{grffile}
\usepackage{caption}
\usepackage{subcaption}
\usepackage{float}
\newcommand{\ket}[1]{|#1\rangle}

\newcommand{\ima}{\mathbbmtt{i}}
\newcommand{\vn}{\mathrm{v}}
\newcommand{\be}{\begin{equation}}
\newcommand{\ee}{\end{equation}}

\begin{document}
\title{Discrete spectrum of the quantum Reissner - Nordstr\"om geometry}
\author{N. Dimakis}
\email{nsdimakis@gmail.com}
\affiliation{Instituto de Ciencias F\'{\i}sicas y Matem\'{a}ticas, Universidad Austral de
Chile, 5090000 Valdivia, Chile}
\author{A. Karagiorgos}
\email{alexkarag@phys.uoa.gr}
\affiliation{Nuclear and Particle Physics Section, Physics Department, University of Athens, GR 157--71 Athens, Greece}
\author{T. Pailas}
\email{teopailas879@hotmail.com}
\affiliation{Nuclear and Particle Physics Section, Physics Department, University of Athens, GR 157--71 Athens, Greece}
\author{Petros A. Terzis}
\email{pterzis@phys.uoa.gr}
\affiliation{Nuclear and Particle Physics Section, Physics Department, University of Athens, GR 157--71 Athens, Greece}
\author{T. Christodoulakis}
\email{tchris@phys.uoa.gr}
\affiliation{Nuclear and Particle Physics Section, Physics Department, University of Athens, GR 157--71 Athens, Greece}

\begin{abstract}
We start from a static, spherically symmetric space-time in the presence of an electrostatic field and construct the mini-superspace Lagrangian that reproduces the well known Reissner - Nordstr\"om solution. We identify the classical integrals of motion that are to be mapped to quantum observables and which are associated with the mass and charge. Their eigenvalue equations are used as supplementary conditions to the Wheeler-DeWitt equation and a link is provided between the existence of an horizon and to whether the spectrum of the observables is fully discrete or not. For each case we provide an orthonormal basis of states as emerges through the process of canonical quantization.
\end{abstract}

\maketitle
\numberwithin{equation}{section}

\section{Introduction}

After the Hamiltonian formulation of General Relativity presented by Dirac \cite{DiracHam}, the canonical formalism introduced by Arnowitt, Deser and Misner \cite{ADM} and the seminal work of DeWitt \cite{DeWitt}, the quantization of gravity has become the holy grail of theoretical physics. In order to surpass several problems that are encountered in the quantization process, other gravitational theories (e.g. Horava-Lifshitz theory \cite{Horava}) or even different approaches like Loop Quantum Gravity (for a general introduction see \cite{Loop1},\cite{Loop2} and references therein) have been proposed. However, regardless of the progress in each theoretical framework, there are still open issues to be addressed \cite{Kiefer}. Hence, one may look for certain simplification schemes that allow the several different paths to be tested on a theoretical level or even to be compared with each other.

In that respect, the mini-superspace approach has been put in use in various cases, so as to simulate in a simple way the quantum behaviour of certain gravitational systems possessing a high degree of symmetry \cite{Bojo,Jala,Asht1,Ewing,Vak2,Bojo2,Barv,Vak1,Pal1,Pal2}. When this reduction takes place, the ensuing configuration is described by a finite number of degrees of freedom and many fundamental difficulties encountered in the quantization of full gravity are, to a large extent, circumvented; at the same time some key distinguishing properties, such as time reparametrization invariance, particular space coordinate covariance \cite{chraut}, existence of constraints, are maintained giving rise to the hope that some properties of the full quantum gravity can be seen by quantizing these reduced systems.  At this level, two main procedures can be followed: the standard canonical quantum mechanics or the polymer quantization \cite{Corichi}, which has been put in use in the framework of Loop Quantum Cosmology.

In the context of the standard canonical quantization, a process involving the classical symmetries of constrained systems \cite{tchrisJP} that are being promoted to operators and used as supplementary conditions to the Wheeler-DeWitt equation has been proposed \cite{tchrsSch}. In that way, quantum observables and their eigenvalues can be related to classical constants of integration appearing in the metric. The method has been implemented in various cosmological configurations \cite{tchriscosmo} as well as black holes \cite{tchrsSch,tchrisRN,tchrisBTZ}, where it can be seen that the implementation of certain allowable subalgebras in the quantization can even lead to a semiclassical avoidance of curvature singularities.

In this work, we revisit the quantization procedure initially presented in \cite{tchrisRN}. We focus our analysis on the quantization with respect to Abelian subalgebras of the symmetries that were not considered there. The procedure followed here leads to an association of the quantum configuration, stemming from the mini-superspace analysis, to the well known P\"oschl - Teller problem of quantum mechanics. Under certain conditions, there exist bound states that lead to a discrete spectrum for the two essential constants appearing in the Reissner-Nordstr\"{o}m metric, the charge $Q$ and the mass $M$. What is more, the appearance of a discrete or a continuous spectrum is seen to be linked with the existence of the horizons, thus in a way seems to be related to the cosmic censorship conjecture \cite{Penrose} (for possible gravitational lensing tests on the cosmic censorship hypothesis see \cite{Ellis}).

The fact that a discrete spectrum appears is highly non trivial, since canonical quantization of mini-superspace gravitational systems usually results in continuous spectra for the observables. Apart from the discrete case - where a Hilbert space can be formally constructed  - we also study the continuous spectrum and provide an orthonormal relation for the corresponding states in terms of the Dirac delta function. Hence, even in that case the quantization can be performed formally in terms of a rigged Hilbert space.

The structure of the paper is the following: in section 2 we present the mini--superspace description of spatially homogeneous and static--spherically symmetric geometries along with the proposed canonical quantization procedure, i.e. the use of the constant potential formulation in the Lagrangian and the promotion of the classical integrals of motion into Hermitian operators. In section 3, the general procedure is applied to a static spherical symmetric spacetime which is controlled by the presence of an electric field: firstly the classical system is analysed in the framework of Lagrangian dynamics and the Reissner--Nordostr\"om spacetime is easily reproduced using the integrals of motion derived through the Killing vector fields of the flat supermetric. Then the quantization of the mini--superspace is carried out in two separate cases; using either a regular or a hyperbolic rotation. Finally a discussion of the results in presented in section 4.

\section{Mini-superspace description and a proposed canonical quantization procedure} \label{section1}

Let us consider the case of Einstein's gravity
\begin{equation} \label{generalact}
  S= \frac{c^3}{16\pi G} \int\!\! \sqrt{-g}\, R\, d^4 x + S_m ,
\end{equation}
where $g$ is the determinant of the space-time metric $g_{\mu\nu}$, $R$ the Ricci scalar and $S_m$ the action of the matter content. For specific types of manifolds possessing a certain group of isometries (e. g. spatially homogeneous or static-spherically symmetric geometries) the variables in the line element can be decoupled in the following manner
\begin{equation}\label{generalline}
  ds^2 =  \epsilon N(x)^2 d x^2 + \gamma_{\kappa\lambda}(x) \sigma^{\kappa}_i(y) \sigma^{\lambda}_j(y) dy^{i} dy^{j},
\end{equation}
where $N$ is the lapse function and $\sigma^{\kappa}_i(y)$'s the invariant basis one-forms
associated to the assumed symmetry group of motions; when $\epsilon=-1$, $x$ is the time variable and $\gamma_{\kappa\lambda}(x)$ the components of the (positive definite) scale factor matrix, while for $\epsilon=1$, $x$ stands for the radial coordinate of some spherically symmetric line element. In these cases, Einstein's equations
\begin{equation}\label{generalEin}
  R_{\mu\nu} - \frac{1}{2} g_{\mu\nu} R = \frac{8\pi G}{c^4} T_{\mu\nu}
\end{equation}
with $T_{\mu\nu} = \frac{2}{\sqrt{-g}} \frac{\delta S_m}{\delta g^{\mu\nu}}$, are reduced to a set of ordinary differential equations with $x$ as the independent dynamical variable. Additionally, if ansatz \eqref{generalline} is inserted into action \eqref{generalact} - and the non-dynamical degrees of freedom are integrated out - there remains a reduced action for a mechanical system consisting out of the finite
number of degrees of freedom left over. The corresponding Euler-Lagrange
equations may (e.g. Bianchi Class A cosmological models), or may not (e.g.
Bianchi Class B cosmological models) be equivalent to the reduced
equations of motion i.e. equations \eqref{generalEin} restricted by hypothesis \eqref{generalline}.
Whenever these two sets are indeed equivalent, we obtain what is called a valid mini-superspace description, and the evolution of the full gravitational system is successfully described by that of the reduced; this
property is a prerequisite for any subsequent quantum treatment of the
reduced action, for in the opposite case we would be quantizing degrees of
freedom whose classical dynamics is not the correct one dictated by
\eqref{generalEin} and \eqref{generalline}. Of course, the crucial question concerning the relation between
any quantum results from this truncated system and the reduction of the
full quantum gravity, can not be answered until this full theory is
constructed. Nevertheless, the fact that properties like time
re-parametrization, existence of constrains, etc. are present in the
reduced system, justifies the hope that its quantization may bare
similarities to the results obtained from a reduction of a full quantum
gravity theory.

The Lagrangians of mini-superspace systems emanating from this procedure assume the general form
\begin{equation}\label{generalLag}
  L= \frac{1}{2 N(x)} \overline{G}_{\alpha\beta}(q) \dot{q}^{\alpha}(x) \dot{q}^{\beta}(x) - N(x) V(q)
\end{equation}
where $\dot{ } = \frac{d}{dx}$. The $q^{\alpha}(x)$'s is a set of variables that incorporates the $\gamma_{\kappa\lambda}$'s plus any matter degree of freedom we consider in $S_m$ (as long as its contribution is quadratic in velocities). The function $V(q)$ and the matrix $\overline{G}_{\alpha\beta}(q)$ are the mini-superspace potential and metric respectively. What is more, Lagrangian \eqref{generalLag} is constrained with the consequence that not all equations of motion are independent from one other. With the help of the Dirac-Bergmann algorithm for singular systems (\cite{Dirac}, \cite{AndBer}) the corresponding Hamiltonian is written as
\begin{equation*}
  H_T = N \mathcal{H} + u_N p_N
\end{equation*}
where
\begin{subequations}
\begin{align}
  p_N & \approx 0 \\ \label{Hamcon}
  \mathcal{H} =\frac{1}{2} \overline{G}^{\alpha\beta} p_\alpha p_\beta + V(q)  & \approx 0
\end{align}
\end{subequations}
are the primary and secondary constraints respectively. The $\approx$ symbol is used to denote a weak equality: A relation which holds on the constraint surface and thus can be used only after all Poisson brackets have been calculated.

For the particular situation at hand, it can be shown that conditional symmetries \cite{Kuchar}, i.e. conserved modulo the constraint \eqref{Hamcon} quantities which are at most linear in the momenta, assume the general form \cite{tchrisJP}, \cite{Dim1}
\begin{equation}\label{genintofmo}
  Q= \xi^{\alpha}(q) p_\alpha + \int\!\! N(x) \left[\omega(q(x))+F(q(x)) \right] V(q(x)) dx
\end{equation}
with
\begin{equation*}
  \mathcal{L}_\xi \overline{G}_{\alpha\beta} = \omega(q) \overline{G}_{\alpha\beta}, \quad \text{and} \quad F(q) =\frac{1}{V(q)} \mathcal{L}_\xi V(q),
\end{equation*}
$\mathcal{L}_\xi$ denoting the Lie derivative with respect to the configuration space vector $\xi$. It is easy to check that
\begin{equation} \label{timeder}
  \frac{d Q}{d x} = \frac{\partial Q}{\partial x} + \{Q,H_T\}= N\, \omega\, \mathcal{H} \approx 0
\end{equation}
holds. Thus, any conformal Killing vector of the mini-supermetric generates a conserved quantity on the constrained surface: When it happens that the conformal factor of $\xi$ over the potential is opposite to the one attained over the metric, i.e. $F=-\omega$ we get
\begin{equation} \label{intofmo}
Q=\xi^\alpha p_\alpha,
\end{equation}
otherwise $Q$ assumes a non-local expression as given in \eqref{genintofmo}.

It is of particular use, especially in quantization \cite{tchrsSch}, to adopt a parametrization that incorporates all the information about the system inside the mini-superspace metric. This can be done by adopting a scaling transformation of the form $N \mapsto n = N\, V$, which allows us to write the equivalent Lagrangian
\begin{equation}\label{generalLag2}
  L= \frac{1}{2 n(x)} G_{\alpha\beta}(q) \dot{q}^{\alpha}(x) \dot{q}^{\beta}(x) - n(x),
\end{equation}
with $G_{\alpha\beta} = V\, \overline{G}_{\alpha\beta}$ being the new, scaled by the potential, mini-superspace metric. The corresponding Hamiltonian constraint becomes
\begin{equation*}
  \mathcal{H} =\frac{1}{2} G^{\alpha\beta} p_\alpha p_\beta + 1   \approx 0
\end{equation*}
and relation \eqref{genintofmo} is still valid for all conformal Killing vectors of this new, scaled by the potential, mini-superspace metric $G_{\alpha\beta}$ by just setting $F=0$ and $N=n/V$. Conserved charges of the form \eqref{intofmo} correspond now to Killing vector fields of $G_{\alpha\beta}$ and have the property of strongly commuting with the Hamiltonian, not just weakly. The latter is evident by \eqref{timeder}, since for the Killing vectors $\omega=0$. This property is extremely useful in the process of quantization, as we shall observe in the following analysis.

Let us proceed by constructing a canonical quantization scheme for system \eqref{generalLag2}. We assume that the mini-superspace $G_{\alpha\beta}$ possesses some Killing vector fields $\xi_I$, where $I$ is an index used to label each one of them. As we discussed, and since we are in the constant potential parametrization, there exist classical integrals of motion of the form \eqref{intofmo} corresponding to each Killing vector $\xi_I$. We follow the usual prescription of assigning differential operators to momenta,
\begin{equation*}
  p_n \mapsto \widehat{p}_n = -\ima\,\hbar \frac{\partial}{\partial n}, \quad p_\alpha \mapsto \widehat{p}_\alpha = - \ima \,\hbar \frac{\partial}{\partial q^\alpha},
\end{equation*}
while the positions are considered to act multiplicatively. In order to address the factor ordering problem of the Kinetic term of $\mathcal{H}$, we choose the conformal Laplacian (or Yamabe operator),
\begin{equation}\label{Hamoperator}
  \widehat{\mathcal{H}} = -\frac{\hbar^2}{2 \mu} \partial_\alpha \left(\mu G^{\alpha\beta}\partial_\beta \right)+ \frac{d-2}{8(d-1)} \mathcal{R} + 1,
\end{equation}
where $\mu(q)= \sqrt{|\det{G_{\alpha\beta}}|}$, $\partial_\alpha = \frac{\partial}{\partial q^\alpha}$, $\mathcal{R}$ the Ricci scalar of the mini-superspace and $d$ its dimension. This choice is uniquely determined by requiring the sought operator to be scalar and covariant under rescalings of the minisuperspace metric; both of these properties hold for the classical system. A very important, as well as interesting, consequence of this choise is the fact that, classical symmetries \eqref{intofmo} are naturally carried over to the quantum description by just assigning to $Q$ the general expression for linear first order, Hermitian operators
\begin{equation}\label{firstordop}
  \widehat{Q}_I = -\frac{\ima \hbar}{2\mu} \left( \mu \xi_I^{\alpha} \partial_\alpha + \partial_\alpha (\mu \xi_I^{\alpha})\right) = -\ima\, \hbar\, \xi_I^\alpha \partial_\alpha
\end{equation}
with the last equality holding due to the $\xi_I$'s being Killing vector fields and $\mu(q)$ the physical measure. Note that only in the constant potential parametrization these symmetries exactly commute with the Hamiltonian (in other parametrizations where the effective potential is not constant they give rise to a multiple of the constraint). Now this property is carried over at the quantum level and
\begin{equation} \label{comQH}
  [\widehat{Q}_I,\widehat{\mathcal{H}}] = 0
\end{equation}
holds. What is more, the classical Poisson algebra of the $Q_I$'s is isomorphic to the quantum algebra of the operators
\begin{equation*}
  \{Q_I,Q_J\} = C^K_{IJ} Q_K \longmapsto [\widehat{Q}_I,\widehat{Q}_J] = -\ima\, \hbar\, C^K_{IJ} \widehat{Q}_K,
\end{equation*}
a fact that is also true for the most general expression in \eqref{firstordop}.

By having constructed a quantization procedure where \eqref{comQH} holds we are able to use the $\widehat{Q}_I$'s as quantum observables together with $\widehat{\mathcal{H}}$. The number of eigenequations
\begin{equation}\label{eigeneq}
  \widehat{Q}_I \Psi = \kappa_I \Psi
\end{equation}
that can be consistently imposed on the wave function is dictated by the integrability condition \cite{tchrisBI}
\begin{equation*}
  C^M_{IJ} \kappa_M = 0
\end{equation*}
where $\kappa_M$ are the eigenvalues and $C^M_{IJ}$ the structure constants of the subalgebra under consideration. The eigen-equations \eqref{eigeneq} are of course used as supplementary conditions together with the primary constraint and the Wheeler-DeWitt equation
\begin{subequations} \label{quantcon}
\begin{align}
  \widehat{p}_n\Psi & = 0 \\ \label{WDW}
  \widehat{\mathcal{H}} \Psi & =0.
\end{align}
\end{subequations}
The latter being imposed according to Dirac's prescription for the quantization of constrained systems, that requires the constraints to annihilate the wave function.

\section{The static, spherically symmetric reduced system}

In this section we study the canonical quantization of the three dimensional flat mini-superspace that arises for the static, spherically symmetric space-times in the presence of an electric field. We start by giving a brief description of the classical system and its conserved quantities before we proceed and use them as quantum observables in the subsequent quantization.

\subsection{Classical Description}
Let us take as our starting point the action \eqref{generalact} where the matter content is
\begin{equation*}
  S_m = -\frac{1}{4 \mu_0}\int\!\! \sqrt{-g}  F^{\mu\nu}F_{\mu\nu} d^4 x
\end{equation*}
with $F_{\mu\nu} = \partial_\mu A_\nu - \partial_\nu A_\mu$. Of course, variation with respect to the metric yields the Einstein-Maxwell set of equations \eqref{generalEin} where the energy - momentum tensor reads
\begin{equation*}
  T_{\mu\nu} = \frac{1}{\mu_0} \left(F_{\mu\lambda}F_{\nu}^{\phantom{\nu}\lambda}- \frac{1}{4} F^{\kappa\lambda}F_{\kappa\lambda} g_{\mu\nu}\right).
\end{equation*}
On the other hand, variation with respect to the vector potential $A^{\mu}$ leads to the vacuum Maxwell equations in the absence of sources
\begin{equation}\label{Maxwell}
  F^{\mu\nu}_{\phantom{\mu\nu};\nu} = 0 .
\end{equation}
From this point on, and for the sake of simplicity, we choose to work in units where $c=4 \pi G= \mu_0=\hbar=1$.

We consider the generic line element
  \begin{equation} \label{statmet}
    ds^2  = - b(r)^2 dt^2 + N(r)^2 dr^2 + a(r)^2 \left(d\theta^2 + \sin^2\theta d\phi^2\right)
  \end{equation}
corresponding to a static, spherically symmetric spacetime in standard coordinates.
The accompanying  form for the electromagnetic potential is
\begin{equation} \label{elpot1}
  A  = f(r) dt .
\end{equation}
Substitution of \eqref{statmet} and \eqref{elpot1} into \eqref{generalact} leads to the following mini-superspace Lagrangian for the configuration at hand
\begin{equation}\label{initLag}
  L = \frac{2}{N(r)}\left(2 a(r) \dot{a}(r)\dot{b}(r) + b(r) \dot{a}(r)^2 + \frac{a(r)^2 \dot{f}(r)^2}{b(r)}\right) + 2 b(r) N(r)
\end{equation}
 It can be easily verified that the Euler-Lagrange equations of \eqref{initLag} are equivalent to the set of Einstein \eqref{generalEin} plus Maxwell \eqref{Maxwell} when reduced by using \eqref{statmet} and \eqref{elpot1}.

As we discussed in the previous section, we shall work in the constant potential parametrization, by setting
\begin{equation*}
  N(r) = \frac{n(r)}{2\, b(r)}.
\end{equation*}
Thus, \eqref{initLag} is transformed into
\begin{equation}\label{initLag2}
  L = \frac{4}{n(r)}\left(2 a(r) b(r) \dot{a}(r) \dot{b}(r)+ b(r)^2 \dot{a}(r)^2+ a(r)^2 \dot{f}(r)^2\right) + n(r)
\end{equation}
from which we deduce the mini-superspace metric
\begin{equation}\label{minimet1}
  G_{\alpha\beta} = \begin{pmatrix}
                      8 b^2 & 8 a b & 0 \\
                      8a b & 0 & 0 \\
                      0 & 0 & 8 a^2
                    \end{pmatrix} ,
\end{equation}
which is just a representation of the three dimensional flat space with Lorentzian signature. Thus, we know that the isometry group is six dimensional and it consists of the three translations and the elements of the $SO(2,1)$ group, i.e. two pseudo-rotations and an actual rotation. In these coordinates we choose to write the six Killing vectors producing autonomous integrals of motion of the form \eqref{intofmo} as
\begin{equation}\label{Kill1}
  \begin{split}
    \xi_1 =& \partial_f, \quad \xi_2 = \frac{1}{a b} \partial_b, \quad \xi_3 = \frac{f}{a b} \partial_b + \frac{1}{a} \partial_f, \quad \xi_4 = -a\partial_a +b\partial_b + f\partial_f \\
    & \xi_5 = \partial_a - \frac{b^2+f^2}{2 a b} \partial_b - \frac{f}{a} \partial_f, \quad \xi_6 = a f\partial_a -b f\partial_b -\frac{b^2+f^2}{2} \partial_f .
  \end{split}
\end{equation}
Additionally, there exists a homothecy $\xi_h = \frac{1}{4}\left(a\partial_a+ b\partial_b + f\partial_f\right)$ giving rise to a non-local conserved quantity of the type \eqref{genintofmo}
\begin{equation}\label{rheon1}
  Q_h = \xi_h^\alpha p_\alpha - \int\!\! n(r) dr .
\end{equation}
In \cite{tchrisRN} it was shown how the space-time solution can be derived algebraically from the set of equations that the integrals of motion provide: First you solve the system $Q_I=c_I$, $I=1,...,5$ and $Q_h =c_h$ ($c_I$, $c_h$ being constants), with respect to $a$, $b$, $\int\!\! n dr$ and their derivatives. Substitution of  this solution in the 6th equation $Q_6=c_6$ just defines the value of the constant $c_6$ in terms of the others, in particular $c_6=-(c_1 c_5+c_3 c_4)/c_2$. Subsequently, the consistency conditions $a' = \frac{da}{dr}$ and $b' = \frac{db}{dr}$ are identically satisfied, while $n = \frac{d}{dr} \int\!\! n dr$ yields the constraint equation that now has reduced to a relation among constants, $c_5=\frac{16-c_3^2}{2 c_2}$.

After a few re-parametrizations: $c_1=4 Q$, $c_2=4/\tilde{c}$, $c_4 = \tilde{c}\, (c_3 Q-4 M)$ and a constant scaling of the $t$ variable with $\pm\tilde{c}$, we arrive at
  \begin{equation} \label{twomet2}
    ds^2  =  - \left(1-\frac{2 M}{a}+ \frac{Q^2}{a^2}\right) dt^2 +\left(1-\frac{2 M}{a}+ \frac{Q^2}{a^2}\right)^{-1} da^2 + a^2 \left(d\theta^2 + \sin^2\theta d\phi^2\right)
  \end{equation}
Of course this is the Reissner-Nordstr\"{o}m metric (\cite{Rpaper}, \cite{Npaper}), with the corresponding electromagnetic potential being
\begin{equation} \label{potentialAsol}
A = \pm \left(\frac{c_3}{4} - \frac{Q}{a(r)}\right) dt
\end{equation}
with the sign depending on how we scaled time to absorb $\tilde{c}$. The constant $c_3$ does not appear in line element \eqref{twomet2}, while in \eqref{potentialAsol} it just defines the value of $A_\mu$ as $a\rightarrow \infty$, hence without loss of generality we can set it equal to zero. Henceforth, we are finally led to
\begin{equation} \label{potentialAsol2}
A = \pm \frac{Q}{a(r)} dt
\end{equation}
where the remaining constant $Q$ appearing here assumes a physical meaning and is to be understood as the absolute value of the charge.

\subsection{Quantization on the flat mini-superspace}

Since \eqref{minimet1} describes the flat space, we can make the coordinate transformation $(a,b,f)\mapsto(\chi,\psi,\zeta)$ with
\begin{equation}\label{metflat1}
  a = \frac{1}{8}\left(\chi - \zeta \right), \quad b =\frac{2 \sqrt{2} \sqrt{\zeta ^2+\psi ^2-\chi ^2}}{\chi -\zeta }, \quad f=\frac{2 \sqrt{2} \psi }{\chi -\zeta }
\end{equation}
to bring the mini-superspace metric into the form $G_{\mu\nu}=\mathrm{diag}(-1,1,1)$. At this point we can identify the linear combinations of $\xi$'s that give the isometries of $G_{\mu\nu}$ as the base elements of the generators of the translations and the $SO(2,1)$ group. Thus, by choosing
\begin{equation}\label{xitoX}
  \begin{split}
    X_1 = \frac{1}{8}\xi_5 -\frac{1}{2}\xi_2, \quad X_2 = \frac{1}{2\sqrt{2}} \xi_3 , \quad X_3 = -\frac{1}{8}\xi_5 -\frac{1}{2}\xi_2  \\
    X_4 = \xi_4 , \quad X_5 = \sqrt{2}\xi_1 + \frac{1}{2\sqrt{2}}\xi_6, \quad X_6 = \sqrt{2} \xi_1 - \frac{1}{2\sqrt{2}}\xi_6
  \end{split}
\end{equation}
we can see that in these coordinates, $X_4$, $X_5$ correspond to hyperbolic rotations (i.e. Lorentz boosts in the $\chi-\zeta$ and $\chi-\psi$ axes) while $X_6$ is the regular rotation in the $\psi-\zeta$ plane; The rest of the $X_i$'s being the translations with respect to each axis
\begin{equation}\label{Xflat}
  \begin{split}
    X_1 & = \partial_\chi, \quad X_2 =  \partial_\psi, \quad X_3 =\partial_\zeta \\
    X_4 = \zeta \partial_\chi  + \chi & \partial_\zeta, \quad X_5 = \psi \partial_\chi + \chi \partial_\psi, \quad X_6 =  \psi\partial_\zeta - \zeta \partial_\psi .
  \end{split}
\end{equation}
The constant values that the conserved quantities $\tilde{Q}_I = X_I^{\alpha}p_\alpha$ acquire on mass shell are
\begin{equation} \label{clasval0}
  \begin{split}
    &\tilde{Q}_1 =\frac{\tilde{c}(16-c_3^2)}{64}-\frac{2}{\tilde{c}}, \quad \tilde{Q}_2 = \frac{c_3}{2 \sqrt{2}}, \quad \tilde{Q}_3 = \frac{\tilde{c}(c_3^2-16)}{64}-\frac{2}{\tilde{c}}, \quad \tilde{Q}_4 = \tilde{c}\, (c_3 Q-4 M)\\
    & \tilde{Q}_5 = \frac{Q \left(128-\tilde{c}^2(16+c_3^2)\right)+8\, \tilde{c}^2 \, c_3\, M }{16\sqrt{2}}, \quad  \tilde{Q}_6 = \frac{Q \left(128+ \tilde{c}^2(16+c_3^2)\right)-8\, \tilde{c}^2 \, c_3\, M }{16\sqrt{2}}
  \end{split}
\end{equation}
in which, if we consider the previously discussed freedom, we can set $c_3=0$ that indicates a vanishing potential at infinity, while $\tilde{c}$ is a constant which is absorbed by a scaling in the base manifold metric and hence it can be assigned to any other value but zero. We shall discuss more of this constant and its value later on in the analysis. At this point lets just observe that the last three vectors of \eqref{Xflat} are related to conserved charges that are associated with essential constants of line element \eqref{twomet2}; namely the mass $M$ and the absolute value of the charge $Q$.

In \cite{tchrisRN} we considered the quantization with respect to several Abelian subalgebras based on the classical symmetries $\xi$ for the Reissner-Nordstr\"{o}m case. In this paper we study the quantization with respect to different algebras involving the quadratic Casimir invariant of the semi-simple subalgebra spanned by $X_4$, $X_5$ and $X_6$
\begin{equation}\label{Qcas}
  Q_{Cas} =  X_6 \otimes X_6 - X_4 \otimes X_4 - X_5 \otimes X_5
\end{equation}
which were not examined in \cite{tchrisRN}. The integral of motion, to which this invariant corresponds, assumes (under use of \eqref{clasval0}) the value
\begin{equation} \label{Qcascla}
   \tilde{Q}_{Cas} = \tilde{Q}_6^2 - \tilde{Q}_4^2-\tilde{Q}_5^2 =16 \tilde{c}^2 (Q^2 - m^2).
\end{equation}
The idea is to quantize the system with respect to either the regular or one of the hyperbolic rotations and use this invariant as a supplementary condition together with the Wheeler-DeWitt equation.

\subsubsection{The regular rotation $X_6$} \label{quantX6}

It is convenient for our purposes to bring $X_6$ into normal form. This is achieved by the pseudo-spherical transformation $(\chi,\psi,\zeta)\mapsto (u,v,w)$
\begin{equation}\label{transnormX6}
  \chi = u \sinh \vn, \quad \psi = u \cosh\vn \cos w, \quad \zeta = u \cosh\vn \sin w,
\end{equation}
which makes the mini-superspace metric read $G_{\alpha\beta} = \mathrm{diag} (1,-u^2,u^2 \cosh^2\vn)$, while $X_6$ becomes just $\partial_w$. At this point, we use \eqref{firstordop} to write down the linear, first order, Hermitian operators, where the $X$'s are used instead of the $\xi$'s, the measure being in these coordinates $\mu = \sqrt{-G} = u^2 \cosh\vn$. As a result, we get
\begin{align}
  \widehat{Q}_6 &= -\ima \frac{\partial}{\partial w} \\
  \widehat{Q}_{Cas} &= \widehat{Q}_6^2-\widehat{Q}_4^2-\widehat{Q}_5^2 = \frac{1}{\cosh\vn}\frac{\partial}{\partial \vn}\left(\cosh\vn\frac{\partial}{\partial \vn}\right) - \frac{1}{\cosh^2\vn} \frac{\partial^2}{\partial w^2},
\end{align}
while the quadratic constraint operator, being derived from \eqref{Hamoperator}, reads
\begin{equation}\label{wdwop}
  \widehat{H} = -\frac{1}{2u^2}\left[\frac{\partial}{\partial u} \left(u^2 \frac{\partial}{\partial u}\right)- \frac{1}{\cosh\vn}\frac{\partial}{\partial \vn}\left(\cosh\vn\frac{\partial}{\partial \vn}\right)+\frac{1}{\cosh^2\vn} \frac{\partial^2}{\partial w}\right] -1.
\end{equation}
As expected, $\widehat{Q}_6$, $\widehat{Q}_{Cas}$ and $\widehat{H}$ form an Abelian algebra of quantum operators and hence we can proceed by finding the common solution of the eigenvalue equations
\begin{subequations} \label{eqigeneq1}
  \begin{align}\label{eqigeneq1a}
    \widehat{Q}_6 \Psi_{k\ell}(u,v,w) & = k \Psi_{k\ell}(u,v,w) \\ \label{eqigeneq1b}
    \widehat{Q}_{Cas} \Psi_{k\ell}(u,v,w) & = \ell (\ell+1) \Psi_{k\ell}(u,v,w),
  \end{align}
\end{subequations}
together with the Wheeler-DeWitt constraint \eqref{WDW}.

In quantum cosmology the spectrum of the operators under consideration is usually continuous, thus it is rather uncommon to be able to distinguish a discrete set of eigenvalues. However, for this algebra this is not the case and a Hilbert space can be formally constructed. The wave function that satisfies \eqref{eqigeneq1} and \eqref{quantcon} can be split into parts $\Psi_{k\ell} = \psi^{(1)}_{\ell}(u)\psi^{(2)}_{k\ell}(\vn)\psi^{(3)}_k(w)$.

Of course, \eqref{eqigeneq1a}  being fixing $ \psi^{(3)}_k(w)$ to
\begin{equation} \label{qsol1a}
  \psi^{(3)}_k(w) =  C_1 e^{\ima \, k \, w}
\end{equation}
with $C_1$ the normalization constant.

The operator $\widehat{Q}_6$ corresponds to the classical symmetry $X_6$, which as we mentioned is a true rotation in the $\psi-\zeta$ plane of the flat configuration space as expressed in the $(\chi,\psi, \zeta)$ variables. Additionally, we see that $w$ appears as the argument of exclusively  trigonometric functions (see transformation \eqref{transnormX6}); thus, in order to cover the hole space exactly once, it can be assumed to attain values in the region $[0,2\pi]$. It is therefore reasonable to assume the boundary condition $\psi^{(3)}_k(0)=\psi^{(3)}_k(2\pi)$, which as, it is well known, results to $k \in \mathbb{Z}$ and $C_1 =(2\pi)^{-1/2}$ so that $\psi^{(3)}_k(w)$ satisfies the normalization condition
\begin{equation*}
  \int_{0}^{2\pi}\psi^{(3)}_{k'}(w)^*\psi^{(3)}_k(w) dw =C_1^{*} C_1 \int_{0}^{2\pi}e^{\ima (k - k')} dw= \delta_{kk'}, \quad k,k' \in \mathbb{Z}
\end{equation*}
where $\delta_{kk'}$ is the Kronecker delta. We proceed to \eqref{eqigeneq1b}, which can be easily seen that it reduces to
\begin{equation} \label{eqVv}
  \frac{1}{\cosh\vn}\frac{d}{d \vn} \left(\cosh\vn \, \frac{d \psi^{(2)}_{k\ell}(\vn)}{d\vn}\right) - \left[ \ell(\ell+1) - \frac{k^2}{\cosh^2\vn}\right] \psi^{(2)}_{k\ell}(\vn)=0.
\end{equation}
At this point, we have to distinguish two cases with respect to the sign of the eigenvalues of the Casimir invariant operator $\widehat{Q}_{Cas}$.

\begin{itemize}
\item Let us first consider $\ell (\ell+1)\geq0$, a condition that holds for $\ell \in \mathbb{R}- (-1,0)$. By performing the transformation $\psi^{(2)}_{k\ell}(\vn)= \frac{\Phi_{k\ell}(\vn)}{\cosh^{1/2}\vn}$, equation \eqref{eqVv} simply becomes
    \begin{equation}\label{PTHam}
      \frac{d^2 \Phi_{k\ell}(\vn)}{d\vn^2} + \left[\frac{k^2-\frac{1}{4}}{\cosh^2\vn} -\frac{1}{4} (2 \ell+1)^2 \right]\Phi_{k\ell}(\vn)= 0.
    \end{equation}
    In \eqref{PTHam} we recognize the one dimensional, time independent Schr\"odinger equation for the famous P\"oschl - Teller potential \cite{Landau}, where we can identify $\frac{1}{4}-k^2 = - V_0<0$ as the depth of the potential well and $E = -\frac{1}{4} (2 \ell+1)^2$ the negative energy that leads to a finite number of bound states. A similar treatment which connects the algebra of operators under consideration with the P\"oschl - Teller system can also be found in \cite{Correa}. In accordance with \cite{Landau} we can introduce new parameters $\kappa$ and $\lambda$ as
    \begin{equation}
      \kappa (\kappa +1) = V_0 \Rightarrow \kappa = |k| -\frac{1}{2}, \quad  \lambda = \sqrt{-E} = |\ell+\frac{1}{2}|
    \end{equation}
    and make the change of variable $\vn \mapsto \sigma = \tanh(\vn)$ which results in \eqref{PTHam} becoming
    \begin{equation}
      \frac{d}{d\sigma} \left[(1-\sigma^2) \frac{d \Phi_{\kappa\lambda}(\sigma)}{d\sigma} \right] +\left[\kappa(\kappa+1) + \frac{\lambda^2}{1-\sigma^2} \right]\Phi_{\kappa\lambda}(\sigma) = 0,
    \end{equation}
    where the original domain  $\vn \in (-\infty,+\infty)$ is now compactified in $\sigma \in (-1,1)$. The  solutions that are finite for $\sigma \rightarrow 1$ are given by
    \begin{equation}\label{solPhi}
      \Phi_{\kappa\lambda}(\sigma) = C_2 (1-\sigma^2)^{\lambda/2} {}_2 F_1 (\lambda-\kappa, \lambda+\kappa+1; \lambda+1 ; \frac{1}{2}(1-\sigma)),
    \end{equation}
    with ${}_2 F_1 (a, b;c;z)$ being the Gauss hypergeometric function and $C_2$ a normalization constant. Regularity of the solution at the border $\sigma \rightarrow -1$ leads to the restriction $\kappa - \lambda= n \in \mathbb{N}$, which for $k$ and $\ell$ reads
    \begin{equation} \label{quantumcond}
    |k| - |\ell+\frac{1}{2}|-\frac{1}{2} =n \in \mathbb{N} \Rightarrow \begin{cases}
                                                                         |k| > \ell, & k \in \mathbb{Z},\quad \ell \in \mathbb{N} \\
                                                                         |k| \leq \ell , & k \in \mathbb{Z}, \quad \ell \in \mathbb{Z}_{-}
                                                                       \end{cases} .
    \end{equation}
    Since $\ell = \nu \in \mathbb{N}$ and $\ell = - \nu-1$ produce the same set of eigenvalues $\ell(\ell+1)$, we can see by the form of the eigenfunction \eqref{solPhi} that, without loss of generality, we need only consider $k \in \mathbb{Z}_+$ and $\ell \in \mathbb{N}$ with $k>\ell$, all other possibilities reproducing the same results. The normalization factor $C_2$ is obtained by expressing the hypergeometric function in terms of either the associate Legendre or the Gegenbauer polynomials and it is calculated to be (for details see \cite{Dong})
    \begin{equation} \label{normc2dis}
      C_2 = \sqrt{\frac{\Gamma(\kappa+\lambda+1)\Gamma(\lambda+\frac{1}{2})}{\pi^{1/2} (\kappa-\lambda)! \Gamma(2\lambda+1) \Gamma(\lambda)}} .
    \end{equation}
    We observe that the quantum conditions for $\ell$ truly satisfy $\ell(\ell+1)\geq 0$, which is the reason why we distinguished this case. If we wish to uncover the classical origin of this inequality we need to calculate the constant value of the integral of motion, given in \eqref{Qcas}, corresponding to the operator $\widehat{Q}_{Cas}$. As we observe from \eqref{Qcascla}, a positive or equal to zero eigenvalue $\ell(\ell+1)$ for $\widehat{Q}_{Cas}$, classically corresponds to a situation where $Q \geq M$. Which, apart for the extremal case $Q= M$, corresponds to a space-time with a naked singularity.

    If we want to associate the classical constants $M$, $Q$ to the quantum numbers $k$ and $\ell$, we have to invert relations $k=\tilde{Q}_6$ and $\ell(\ell+1)= Q_{Cas}$ leading to
    \begin{subequations}\label{chargesqm}
      \begin{align} \label{masstoklc2}
         M & = \frac{\sqrt{32 \tilde{c}^2 k^2 - (\tilde{c}^2+8)^2 \ell (\ell+1)}}{4\, |\tilde{c}| (\tilde{c}^2+8)} \\
         Q & = \frac{\sqrt{2}\, k}{\left(\tilde{c}^2+8\right)}.
      \end{align}
    \end{subequations}
    where we have already considered $c_3=0$, that is the physical condition for the potential to vanish at infinity. In general, the values of the conserved quantities $\tilde{Q}_I$ depend from the integration constants $\tilde{c},c_3$. This situation reflects the fact that the different choices of these constants merely correspond to alternative representatives  of the underlying geometry of the space time; changing the values of the constants we just change the numerical form of the line element but the geometry remains the same. Thus, there values are at our disposal provided that they are chosen in such a way that respect the physical content of the theory.

    The physical parameters of the problem, are the mass $M$ and the absolute value of the charge $Q$; we thus  need two operators to correspond to these essential constants. From the form of $\tilde{Q}_I$ in \eqref{clasval0}, we observe that the last three are associated with these values and they are formed from the two hyperbolic rotations $\tilde{X}_4,\tilde{X}_5$ along with the regular rotation $\tilde{X}_6$. In order to end up with two operators, that each one distinctively contains the information of $M$ and $Q$, we may eliminate the conserved quantity of one of the hyperbolic rotations, since there are essentially the same. Thus, in addition to the previously discussed freedom of setting $c_3=0$, we must take $\tilde{c}=2\sqrt{2}$, resulting to
    \begin{equation}
      \begin{split}
        &\tilde{Q}_1=0, \quad \tilde{Q}_2=0, \quad \tilde{Q}_3=-\sqrt{2} \\
        & \tilde{Q}_4=-8\sqrt{2}M, \quad \tilde{Q}_5=0, \quad \tilde{Q}_6=8\sqrt{2}Q.
      \end{split}
    \end{equation}
    while at the same time \eqref{chargesqm} become
    \begin{subequations} \label{masschargetoklc3}
      \begin{align} \label{masstoklc3}
         M & = \frac{1}{8\sqrt{2}} \sqrt{k^2-\ell(\ell+1)} \\
         Q & = \frac{k}{8\sqrt{2}}.
      \end{align}
    \end{subequations}
    As we can see, under this choice, \eqref{masstoklc3} implies that the demand for the reality of the classical mass $M$ results, through the above association, in  the quantum condition $k>\ell$ (when $\ell(\ell+1)$ is positive) with $k$ an integer and $\ell$ a natural number.

    Finally, we close the study of this case by giving the solution that satisfies the constraint equation \eqref{WDW}
    \begin{equation} \label{qsol1c}
      \psi^{(1)}_{\ell}(u)=  C_4 j_{\ell}(\sqrt{2} u) + C_5 y_{\ell}(\sqrt{2} u),
    \end{equation}
    where the $j_{\ell}$, $y_{\ell}$ are the spherical Bessel functions of the first and second kind respectively and the $C_4$, $C_5$ complex constants of integration. Due to the fact that the spherical Bessel $j_{\ell}(x)$ is well behaved at zero and infinity, while at the same time the following orthonormality condition holds
    \begin{equation*}
      \int_{0}^{+\infty}\!\! u^2 j_{\ell}(\alpha_1 u) j_{\ell}(\alpha_2 u) du = \frac{\pi}{2 \alpha_1^2} \delta(\alpha_1 - \alpha_2),
    \end{equation*}
    we need only consider $C_4 = \frac{2}{\sqrt{\pi}}$ and $C_5=0$ in \eqref{qsol1c} to be able to write
    \begin{equation}\label{psiorth3}
       \int_{0}^{+\infty}\!\! u^2 \psi^{(1)}_{\ell}(u)^{*} \psi^{(1)}_{\ell}(u) du = C_4^{*}C_4\int_{0}^{+\infty}\!\! j_{\ell}(\sqrt{2} u) j_{\ell}(\sqrt{2} u)du = \delta(0),
    \end{equation}
    where we symbolically choose to express the right hand side of the above equation as $\delta(0)$ with respect to which we are able to normalize the probability density
    \begin{equation}
       \rho(u,\vn,w) = \frac{\mu \Psi_{k\ell}^{*} \Psi_{k\ell}}{\delta(0)},
    \end{equation}
    with the values of the latter for $u \in (0,+\infty)$, $\vn \in \mathbb{R}$ and $w\in [0,2\pi]$ lying  between zero and one.

\item As we already saw, the previous case corresponding to the classical restriction $Q\geq M$ leads to bound states in a P\"oschl - Teller system. In order to study what happens when classically $Q< M$, we need to consider $\ell (\ell+1)<0$. This leads us to assume a complex quantum number $\ell = -1/2 +\ima s$, $s \in \mathbb{R}$. This is not a contradiction with the fact that the operator $\widehat{Q}_{Cas}$ is constructed as a Hermitian operator, because the eigenvalue is not $\ell$ but the combination $\ell(\ell+1)$ which is still real but negative. This time a discrete set of eigenvalues for $\ell$ cannot be derived, since the ``energy" $E = -\frac{1}{4} (2 \ell+1)^2$ of the corresponding system we studied earlier is positive and above the potential well. However, if we convert the solution back to the variables where we have obtained equation \eqref{eqVv} it is written as
    \begin{equation}
      \psi^{(2)}_{k\ell}(\ima \sinh \vn) = C_2 P^k_{\ell}(\ima \sinh \vn) + C_3 Q^k_{\ell}(\ima \sinh \vn) ,
    \end{equation}
    where $P^k_{\ell}(z)$ and $Q^k_{\ell}(z)$ are the associated Legendre functions of the first and second kind respectively.

    We already know from the P\"oschl - Teller system that the spectrum is now continuous. However, an orthogonality relation can still be deduced for both of the functions (for details see appendix \ref{AppA}) and it is of the form
    \begin{equation}\label{psiorth2b}
      \int_{-\infty}^{+\infty}\!\!  (\psi^{(2)}_{k,\ima s'-1/2}(\vn))^{*}  \psi^{(2)}_{k,\ima s-1/2}(\vn) d \sinh\vn \propto \delta(s' - s) +\delta(s'+s), \quad s \in \mathbb{R}.
    \end{equation}
    We have to note that $s$ and $-s$ both correspond to the same eigenvalue $\ell(\ell+1)$, thus explaining the double delta's appearing on the right hand side of \eqref{psiorth2b}. Furthermore, as can also be seen in appendix \ref{AppA}, for function $P_{\ell}^k (\ima\, \sinh\vn)$ the probability amplitude is the same for both of these values, since the multiplying factor is symmetric to the change $s\mapsto -s$. On the contrary, this is not the case if we consider $Q_{\ell}^k (\ima\, \sinh\vn)$ whose relevant expression does not possess this symmetry.

    Finally, the constraint equation $\widehat{\mathcal{H}} \Psi_{k\ell}=0$ is still satisfied by \eqref{qsol1c}. Additionally, relations \eqref{masstoklc3} still hold with $\ell(\ell+1)$ being now always negative and thus $M$ assumes once more only real values.
\end{itemize}

Let us sum the results we obtained: For $\ell$ discrete and $\ell(\ell+1)\geq 0$, the states $\ket{k,\ell}$ are normalized to the infinity of the $\delta(0)$ appearing in \eqref{psiorth3} and the wave function reads
\begin{equation} \label{discretepsi}
  \Psi_{k\ell} = \frac{\sqrt{2}}{\pi} C_2(k,\ell)\,  j_{\ell}(\sqrt{2} u)\, \frac{{}_2 F_1 (\ell-k+1, \ell+k+1; \ell+\frac{3}{2} ; \frac{1}{2}(1-\tanh\vn))}{\cosh^{\ell-1/2}\vn} e^{\ima k w},
\end{equation}
with the constraints $k \in \mathbb{Z}^{+}$, $\ell \in \mathbb{N}$ and $k> \ell$. The constant $C_2(k,\ell)$ is given by \eqref{normc2dis} under the substitution of $\kappa=k-1/2$, $\lambda = \ell+1/2$. On the other hand, for the continuum case $\ell(\ell+1)< 0$, that classically corresponds to $M>Q$, we get: for $\ell=-1/2+\ima \, s$ and by considering \eqref{ApporthcontP} and \eqref{coefAofP} a wave function of the form
\begin{equation}
  \Psi_{k s} = \frac{1}{\pi} \left(\frac{\Gamma(\frac{1}{2}-k-\ima\, s)\Gamma(\frac{1}{2}-k+\ima\, s)}{\cosh(s\pi)\Gamma(-\ima\, s)\Gamma(\ima\, s)} \right)^{1/2} j_{\ima\, s-\frac{1}{2}}(\sqrt{2} u)\, P^k_{\ima\, s-\frac{1}{2}} (\ima\,\sinh\vn)\,e^{\ima k w}
\end{equation}
with $k \in \mathbb{Z}$ and $s \in \mathbb{R}$, normalized to the product of $\delta(0)$ with the deltas of \eqref{psiorth2b}. We remind that in any case the measure function is $\mu = u^2 \cosh \vn$, while the domain of the variables is $u \in (0,+\infty)$, $\vn \in \mathbb{R}$ and $w \in [0,2\pi]$.

Additionally, for the discrete case, one can define ladder operators as
\begin{subequations}
  \begin{align}
    \widehat{A}_+ = \widehat{Q}_4 - \ima \widehat{Q}_5 \\
    \widehat{A}_- = \widehat{Q}_4 + \ima \widehat{Q}_5
  \end{align}
\end{subequations}
that raise or lower the state with respect to the eigenvalue $k$. The usual algebra between these operators and $\widehat{Q}_6$ is satisfied
\begin{equation}
  [\widehat{A}_+,\widehat{A}_-] = -2 \widehat{Q}_6, \quad [\widehat{Q}_6,\widehat{A}_+] = \widehat{A}_+, \quad [\widehat{Q}_6,\widehat{A}_-] = -\widehat{A}_-
\end{equation}
and of course the Casimir invariant can be rewritten in terms of $\widehat{A}_{\pm}$ and $\widehat{Q}_6$ as
\begin{equation}
  \widehat{Q}_{Cas} = \widehat{Q}_6 \left(\widehat{Q}_6+1\right) - \widehat{A}_-\widehat{A}_+ .
\end{equation}
It is straightforward to check that, by using \eqref{discretepsi} and recurrence relations that connect hypergeometric functions of successive indexes, the following relations are satisfied
\begin{subequations}
  \begin{align}
    \widehat{A}_+ \ket{k,\ell} &= \left[k(k+1)-\ell(\ell+1)\right]^{1/2} \ket{k+1,\ell} \\
    \widehat{A}_- \ket{k,\ell} &= \left[k(k-1)-\ell(\ell+1)\right]^{1/2} \ket{k-1,\ell} .
  \end{align}
\end{subequations}
with the action of the annihilation operator on the lowest state for $k$ being of course zero, $\widehat{A}_- \ket{\ell+1,\ell}=0$.

\subsubsection{The hyperbolic rotation $X_4$}

There is no difference in treating the two cases where the basic eigenoperator is one of the two hyperbolic rotations. Thus, without loss of generality we solely consider quantization with respect to $X_4$. As we previously did for the regular rotation, we choose to bring the generator into normal form by performing the transformation
\begin{equation}\label{transnormX4}
  \chi = u \sinh \vn \cosh w , \quad \psi = u \cosh\vn, \quad \zeta = u \sinh\vn \sinh w .
\end{equation}
Note that the coordinates $(u,\vn,w)$ are not the same used in section \ref{quantX6}. Under \eqref{transnormX4}, the mini-superspace metric of the flat space becomes $G_{\alpha\beta} = \mathrm{diag} (1,-u^2,u^2 \sinh^2\vn)$, leading to a measure function $\mu = \sqrt{-G} = u^2 |\sinh\vn|$. In these coordinates, the operators that form an Abelian subalgebra are
\begin{align}
  \widehat{Q}_4 &= -\ima \frac{\partial}{\partial w} \\
  \widehat{Q}_{cas} &= \widehat{Q}_6^2-\widehat{Q}_4^2-\widehat{Q}_5^2 = \frac{1}{\sinh\vn}\frac{\partial}{\partial \vn}\left(\sinh\vn\frac{\partial}{\partial \vn}\right) - \frac{1}{\sinh^2\vn} \frac{\partial^2}{\partial w^2} \\
  \widehat{H} &= -\frac{1}{2u^2}\left[\frac{\partial}{\partial u} \left(u^2 \frac{\partial}{\partial u}\right)- \frac{1}{\sinh\vn}\frac{\partial}{\partial \vn}\left(\sinh\vn\frac{\partial}{\partial \vn}\right)+\frac{1}{\sinh^2\vn} \frac{\partial^2}{\partial w}\right] -1.
\end{align}
Once more, a function of the form $\Psi_{k\ell} = \psi^{(1)}_{\ell}(u)\psi^{(2)}_{k\ell}(\vn)\psi^{(3)}_k(w)$ satisfies the set of equations
\begin{equation*}
  \widehat{Q}_4 \Psi_{k\ell} = k \Psi_{k\ell}, \quad \widehat{Q}_{cas} \Psi_{k\ell} = \ell (\ell+1) \Psi_{k\ell}, \quad \widehat{H} \Psi_{k\ell} = 0
\end{equation*}
with
\begin{subequations} \label{quantsol2}
\begin{align} \label{qsol2a}
  \psi^{(3)}_k(w) = & C_1 e^{\ima \, k \, w} \\ \label{qsol2b}
  \psi^{(2)}_{k\ell}(\vn) = & C_2 P_\ell^{\ima\,k} (\cosh\vn) + C_3 Q_\ell^{\ima\, k} (\cosh\vn)\\ \label{qsol2c}
  \psi^{(1)}_{\ell}(u)= & C_4 j_{\ell}(\sqrt{2} u) + C_5 y_{\ell}(\sqrt{2} u) .
\end{align}
\end{subequations}
As we observe, the difference with respect to the previous case examined in \ref{quantX6} - apart from the fact that $(u,\vn,w)$ are connected differently to the original variables $(a,b,f)$ - lies in the form of \eqref{qsol2b}. What is more, due to the fact that $w$ is not a periodic variable it does not lead to a discrete spectrum for $k$, on the contrary the orthogonality condition for \eqref{qsol2a} is now
\begin{equation*}
  \int_{-\infty}^{+\infty}\!\! \psi^{(3)}_{k'}(w)^{*} \psi^{(3)}_k(w) dw = C_1^{*}C_1 \int_{-\infty}^{+\infty}\!\! e^{-\ima(k' - k)w } dw = 2\pi C_1^{*}C_1  \delta (k'-k),
\end{equation*}
from which we deduce again that $C_1 = (2\pi)^{-1/2}$.

The situation regarding $\psi^{(2)}_{k\ell}(\vn)$ expressed by solution \eqref{qsol2b} is not as clear as for the previous algebra and an orthonormality relation with respect to $\ell$ cannot be produced in a similar manner. However, for both $P_\ell^{\ima\,k} (x)$ and $Q_\ell^{\ima\,k} (x)$, as can be seen by the expansions \eqref{asymptoticP} and \eqref{asymptoticQ}, there are values of $\ell$ for which they satisfy the boundary conditions set in appendix \ref{AppA}, where $\Psi$ is required to vanish at the boundary. As $\vn \rightarrow \pm \infty$, $x=\cosh\vn \rightarrow +\infty$ the functions $P_\ell^{\ima\,k} (x)$, $Q_\ell^{\ima\,k} (x)$ both tend to zero if $\ell \in (-1,0)$ and for any value of $s\in \mathbb{R}$ if $\ell = -1/2 +\ima s$. Both of these cases however, solely correspond to $\ell (\ell+1)<0$, which in its turn implies the classical relation $Q<M$.

\section{Discussion}

In the present paper we have derived the mini-superspace Lagrangian of a static, spherically symmetric space-time in the presence of an electrostatic field. The solution of this model is the well known Reissner - Nordstr\"om solution.
The classical integrals of motion, were then mapped to quantum observables and their eigenvalue equations were used as supplementary conditions along with the Wheeler-DeWitt equation, so that the wave function is determined up to normalization constants. Through a careful examination of the geometrical and physical properties of the system we have singled out the operators corresponding to mass and charge $\hat Q_6$ and $\hat Q_{Cas}$. The application of reasonable boundary conditions leads: a) to a purely discrete spectrum for the two operators, with an orthonormal basis of state vectors, through the known properties of the P\"oschl-Teller system, when for the classical system $Q\geq M$ holds and b) to discrete for $\hat Q_6$ but continuous for $\hat Q_{Cas}$ in the case when $Q< M$. The discrete spectrum of the naked singularity case could naively be used to model particles, however only if we could consider them as static, spherically symmetric configurations.

As far as the classical singularity is concerned, it is known that there is no general agreement on what could signify its possible avoidance. Several proposals have been made, like the vanishing of the wave function on the singular point (as a boundary condition) or that the probability density or even the probability itself to be zero near that region. The latter route is the one that we choose to follow in this work. In our parametrization, as we can see from the line element \eqref{twomet2}, the spacetime scalar curvatures are $S=R^{\mu\nu}R_{\mu\nu}=\frac{4 Q^4}{a^8}$ and $K=R^{\mu\nu\kappa\lambda}R_{\mu\nu\kappa\lambda}=\frac{8 \left(6 m^2 a^2-12 m Q^2 a+7 Q^4\right)}{a^8}$, indicating that the singularity lies at the plane $a=0$ of the configuration space.

By taking in account transformations \eqref{metflat1} and \eqref{transnormX6} we can see that $a$ appears only in the $u$ variable as $u \propto a\, b$. Hence, in order to demonstrate what happens when $a\rightarrow 0$ we need only check the behaviour of the $u$ dependent part of the probability
\begin{equation} \label{probsing}
  P_\varepsilon = I_{\mathrm{v}w} \int_{0}^{\varepsilon}u^2 j_\ell (\sqrt{2} u) j_{\ell}(\sqrt{2} u)^* du
\end{equation}
at the limit $\varepsilon \rightarrow 0$. With $I_{\mathrm{v}w}$ we denote the rest of the integrals involving the $\mathrm{v}$ and $w$ variables, which are either finite (discrete case) or normalized in terms of delta functions (continuum case).

For the discrete case where $j_\ell (u)$ is real we get
\begin{equation}
  P_\varepsilon = I_{\mathrm{v}w} \left[u^2 \left(  \frac{\pi}{4\sqrt{2}}J_{\ell+1/2}\left(\sqrt{2} u\right)^2-J_{\ell-1/2}\left(\sqrt{2} u\right) J_{\ell+3/2}\left(\sqrt{2} u\right)\right)\right]_{0}^{\varepsilon},
\end{equation}
where with the help of the properties of the Bessel function $J_{\mu}(x)$ it easily derived that $\underset{\varepsilon\rightarrow 0}{\lim} P_\varepsilon =0$. On the other hand, for the continuous case, where $\ell=-1/2+\ima\, s$, integral \eqref{probsing} leads to
\begin{equation}
  P_\varepsilon = I_{\mathrm{v}w} \left[\frac{s\, \sinh(\pi s)}{4\sqrt{2}} \left(\, _1F_2\left(-\frac{1}{2};-\ima\, s,\ima\, s;-2 u^2\right)-1\right)\right]_{0}^{\varepsilon},
\end{equation}
which again becomes zero as $\varepsilon\rightarrow 0$, because $_1F_2\left(-\frac{1}{2};-\ima\, s,\ima\, s;0\right)=1$. So, in both cases we have vanishing of the probability at the singularity, which implies that there is zero
transition probability from a non-singular configuration $(u\neq0)$ to the
singular one $(u=0)$.

Finally, we can close our discussion with a comment on relations \eqref{masschargetoklc3}. If we want to reinstate $\hbar$ in them, the latter should appear normally in their right hand side. However, due to the fact that the quantization is not being performed with $t$ as the dynamical parameter but with the radial distance $r$ instead, it is not that straightforward to consider the usual $\hbar\simeq 1.054 \cdot 10^{-34} \mathrm{J \cdot sec}$ as the correct constant to utilize here. If we use $d$ instead of $\hbar$ for the symbol of this constant (so as to avoid confusion with the real $\hbar$), we must write
\begin{subequations} \label{mQwithd}
  \begin{align}
    M & = \frac{1}{8\sqrt{2}} \sqrt{k^2-\ell(\ell+1)} d \\ \label{Qwithd}
    Q & = \frac{1}{8\sqrt{2}} k \, d .
  \end{align}
\end{subequations}
As it happens, both $M$ and $Q$ appearing in the metric have units of distance. The real physical mass $m_0$ and charge $q$ are related to them through the well known relations
\begin{equation} \label{realmasscharge}
  M =\frac{G \, m_0}{c^2}, \quad Q=\frac{q}{c^2} \sqrt{\frac{G}{4\pi \varepsilon_0}}.
\end{equation}
Due to \eqref{mQwithd} we expect that the constant $d$ should assume the notion of some fundamental distance, in contrast to $\hbar$ whose units are those of angular momentum. Let us take \eqref{Qwithd} and substitute the value that relates it to the real physical charge $q$. Then, for the basic state of the discrete spectrum, $k=1$, we derive
\begin{equation} \label{fundconst}
d= 4\sqrt{2} \frac{q}{c^2} \sqrt{\frac{G}{\pi \varepsilon_0}} .
\end{equation}
By assuming for the charge the lowest possible value in nature, that is $q=|e|\simeq 1.602 \cdot 10^{-19} \mathrm{C}$, while the speed of light, the vacuum permittivity and Newton's gravitational constant are respectively: $c\simeq 2.998 \cdot 10^{8} \mathrm{m/sec}$, $\varepsilon_0 \simeq 8.854 \cdot 10^{-12} \mathrm{C^2/N\cdot m^2}$ and $G\simeq 6.674 \cdot 10^{-11} \mathrm{m^3/kgr\cdot sec^2}$; we get
$d\simeq 1.562 \cdot 10^{-35} \mathrm{m}$ which is very close and of the same magnitude as the Planck length $\ell_P \simeq 1.616 \cdot 10^{-35} \mathrm{m}$. As we see, $M$ and $Q$ are quantized this way in terms of a distance which is similar to that of the Planck length.

From Dirac's quantization condition
\begin{equation}\label{Diraccond}
  \frac{q q_m}{2 \pi \varepsilon_0 \hbar c^2}= n \in \mathbb{Z},
\end{equation}
where $q_m$ is the charge of a hypothetic magnetic monopole, it is easy to derive that the combination $e^2/\varepsilon_0 c$ has units of $\hbar$ and is proportional to it. This can be seen by substitution of $q=|e|$ and $q_m=|e| c$ in \eqref{Diraccond}. Thus, it is not surprising that the combination of fundamental constants appearing in \eqref{fundconst} leads to something that is of the same order as the Planck length. Relations \eqref{Qwithd} and \eqref{Diraccond} can also be combined to derive an expression linking $k$ and $n$. For example if we consider $d\sim \ell_p=\sqrt{\frac{\hbar G}{c^3}}$ and use $Q$ from \eqref{realmasscharge} we can deduce that $k^2=64\, n$. However, such a correspondence may be precarious, mostly due to the fact that the consideration of a monopole charge $q_m$ in our case would totally alter the quantization procedure: Even though that classically a possible inclusion of a monopole charge in $F_{\mu\nu}$ does not alter the Reissner - Nordstr\"om geometry (in place of $Q^2$ in the metric \eqref{twomet2} there would appear the sum of the squares of the two charges), it can be seen that the reduced Lagrangian is different and the corresponding mini-superspace metric is no longer flat. Thus, the previous results regarding the quantum analogues of $M$ and $Q$ cannot remain the same. Unlike the classical solution, the quantization procedure is sensitive to the source of the field.

\appendix

\section{Orthogonality relations for the $\ell(\ell+1)<0$ case.} \label{AppA}

As we discussed in section \ref{quantX6}, the $\ell(\ell+1)>0$ case leads to a pure discrete spectrum for $k$ and $\ell$, when $k \in \mathbb{Z}_{+}$ and $\ell \in \mathbb{N}$ with $k>\ell$. By going back to the classical level, so as to see what constraint the inequality $k>\ell$ imposes on the classical constants, we observe that it corresponds to a situation where $Q \geq M$. Thus, it describes a black hole (for the static, spherically symmetric case) only in the extremal case $Q=M$ and a space-time with a classical naked singularity when $Q>M$. In order to go over to the region $Q <M$ we need to consider the eigenvalue $\ell (\ell+1)$ of the Casimir invariant to be negative. This can happen by allowing for complex quantum numbers of the form $\ima\, s - 1/2$. Hence, we need to study the behaviour of $P^k_{\ima s- 1/2}(\ima\, x)$ and $Q^k_{\ima s- 1/2}(\ima\, x)$.

\subsection{Orthogonality condition for $P^k_{\ima s- 1/2}(\ima\, x)$}

First of all we need to check if the solution satisfies the boundary conditions that are to be imposed so that the operators are Hermitian. In our case this translates to the wave function vanishing at the boundary of the real line $(-\infty,+\infty)$ which is the domain of the variable $x=\sinh\vn$. By using the series representation for $P^k_{\ell}(z)$, that is valid when $|1-z|/2 >1$ and $2 \ell \notin \mathbb{Z}$ (\cite{Abr}-\cite{Wolf}), we are led to
\begin{equation} \label{asymptoticP}
  \begin{split}
    P^k_{\ell}(z) = \frac{1}{\sqrt{\pi}} \frac{(1+z)^{k/2}}{(1-z)^{k/2}} \Bigg[ \frac{2^\ell \Gamma(\ell+\frac{1}{2})}{\Gamma(\ell-k+1)}(z-1)^\ell + \frac{2^{-\ell-1} \Gamma(-\ell-\frac{1}{2})}{\Gamma(-k-\ell)} (z-1)^{-\ell-1} + \\
     O\left(\frac{1}{1-z}\right)\Bigg].
  \end{split}
\end{equation}
We set into the above relation $\ell = \ima\, s-1/2$, $z=\ima\, x$ and consider two possibilities regarding the sign of $x$:
\begin{itemize}
  \item Case $x>> 1$: The approximation $\ima\, x-1 \simeq \ima\, x$, together with the fact that $\frac{(1+\ima\, x)^{k/2}}{(1-\ima\, x)^{k/2}} \rightarrow e^{\ima k \pi/2}$ as $x$ tends to plus infinity leads to
      \begin{equation} \label{Ppos}
        P^k_{\ima s- 1/2}(\ima\, x) \simeq \frac{1}{\sqrt{x}} \left(\alpha_{k,s} x^{\ima s} + \beta_{k,s} x^{-\ima s} \right)
      \end{equation}
      with
      \begin{subequations} \label{abks}
        \begin{align}
          \alpha_{k,s} = & \frac{2^{\ima s}}{\sqrt{2\pi}} e^{\ima (k-\frac{1}{2})\frac{\pi}{2}} \frac{\Gamma (\ima s)}{\Gamma(\frac{1}{2}-k+\ima s)} e^{-s \frac{\pi}{2}} \\
          \beta_{k,s} = & \alpha_{k,-s}
      \end{align}
      \end{subequations}
  \item Case $x<< -1$: By using the fact that $\underset{x\rightarrow -\infty}{\lim}\frac{(1+\ima \, x)^{k/2}}{(1- \ima \, x)^{k/2}} \rightarrow e^{-\ima k \pi/2}$ and again  $\ima\, x-1 \simeq \ima\, x$ we arrive to the expression
      \begin{equation} \label{Pneg}
        P^k_{\ima s- 1/2}(\ima\, x) \simeq \frac{1}{\sqrt{-x}} \left(\zeta_{k,s} (-x)^{\ima s} + \eta_{k,s} (-x)^{-\ima s} \right)
      \end{equation}
      where
      \begin{subequations} \label{zeks}
        \begin{align}
          \zeta_{k,s} = & \frac{2^{\ima s}}{\sqrt{2\pi}} e^{-\ima (k-\frac{1}{2})\frac{\pi}{2}} \frac{\Gamma (\ima s)}{\Gamma(\frac{1}{2}-k+\ima s)} e^{s \frac{\pi}{2}} \\
          \eta_{k,s} = & \zeta_{k,-s}
      \end{align}
      \end{subequations}
\end{itemize}
Thus, as we can see from \eqref{Ppos} and \eqref{Pneg}, $P^k_{\ima s- 1/2}(\ima\, x)$ vanishes at the border of $(-\infty,+\infty)$ satisfying the necessary boundary conditions.

By applying the change of variable $x=\sinh\vn$ in \eqref{eqVv} the latter becomes the associate Legendre equation with solutions the corresponding functions with purely imaginary argument. Let $W^k_\ell(\ima\, x)$ and $W^m_n(\ima\, x)$ be any of the two associated Legendre functions, then equations
\begin{subequations}\label{ortheqs}
  \begin{align}\label{ortheq1}
    & \frac{d}{dx}\left[(1+x^2)\frac{d W^k_\ell}{dx}\right] -\left[\ell(\ell+1)-\frac{k^2}{1+x^2}\right]W^k_\ell =0 \\ \label{ortheq2}
    & \frac{d}{dx}\left[(1+x^2)\frac{d W^m_n}{dx}\right] -\left[n(n+1)-\frac{m^2}{1+x^2}\right]W^m_n =0
  \end{align}
\end{subequations}
identically hold. If we multiply the first with $W^m_n$, the second with $W^k_\ell$ and subtract we get
\begin{equation*}
  \left[(\ell-n)(\ell+n+1)-\frac{k^2-m^2}{1+x^2}\right]W^k_\ell W^m_n = \frac{d}{dx}\left[(1+x^2)\left(W^m_n\frac{d W^k_\ell}{dx}- W^k_\ell \frac{d W^m_n}{dx}\right)\right],
\end{equation*}
which, for $k=m$ and $\ell\neq n$ becomes
\begin{equation} \label{orthcond}
  (\ell+n+1)\int_{-\infty}^{+\infty}\!\! W^k_\ell(\ima\, x) W^m_n(\ima\, x) dx = \left[\frac{1+x^2}{\ell-n}\left(W^m_n\frac{d W^k_\ell}{dx}- W^k_\ell \frac{d W^m_n}{dx}\right)\right]_{-\infty}^{+\infty}.
\end{equation}
In place of $W^k_\ell$ let us consider $P^k_\ell$ and in $W^m_n$ its complex conjugate. If we set $\ell=\ima\, s-1/2$ and $n=\ima\, p-1/2$ with $s, p \in \mathbb{R}$ we can write
\begin{equation} \label{genint1}
  \begin{split}
    \int_{-\infty}^{+\infty}\!\! P^k_{\ima s-1/2}(\ima\, x) & (P^k_{\ima p-1/2}(\ima\, x))^{*} dx = \\ & = \left[\frac{1+x^2}{p^2-s^2}\left((P^k_{\ima p-1/2})^{*}\frac{d P^k_{\ima s-1/2}}{dx}- P^k_{\ima s-1/2} \frac{d (P^k_{\ima p-1/2})^{*}}{dx}\right)\right]_{-\infty}^{+\infty} \\
    & = \Big[\frac{1}{p^2-s^2}\mathcal{A}(x)\Big]_{-\infty}^{+\infty} .
  \end{split}
\end{equation}
Since we are interested in the limit of the above expression in the brackets at $\pm \infty$ we can use the approximate relations \eqref{Ppos} and \eqref{Pneg}. We follow a similar analysis in the spirit of \cite{Nostrand} and \cite{Gotze}. Firstly, for the case when $x>0$ - and by the use of Euler's formula $x^{\ima s} = \cos(s\ln x)+\ima \sin(s\ln x)$ - we deduce that, for $x>>1$, $\mathcal{A}(x)$ becomes
\begin{equation*}
  \begin{split}
    \mathcal{A}(x)\big|_{x>>1}=:\mathcal{A}_{+}(x)  = & \ima (p+s) \cos [(p-s)\ln x] \left(\alpha_{k,p}^{*}\alpha_{k,s} - \beta_{k,p}^{*}\beta_{k,s}\right) \\
    & + \ima (p-s) \cos[(p+s)\ln x] \left(\alpha_{k,p}^{*} \beta_{k,s} - \beta_{k,p}^{*}\alpha_{k,s}\right) \\
    & + (p+s) \sin[(p-s)\ln x] \left(\alpha_{k,p}^{*}\alpha_{k,s} + \beta_{k,p}^{*}\beta_{k,s}\right) \\
    & + (p-s) \sin[(p+s)\ln x] \left(\alpha_{k,p}^{*} \beta_{k,s} + \beta_{k,p}^{*}\alpha_{k,s}\right) .
  \end{split}
\end{equation*}
At this point - and since for the continuous spectrum calculations involving the wave function require integrals with respect to the eigenvalues - we choose to interpret the result of \eqref{genint1} in a distributional way. It is known by the Riemann-Lebesgue lemma that
\begin{equation} \label{RimLebL}
  \underset{y\rightarrow +\infty}{\lim} \int_{\mathbb{R}}\!\! f(s) \cos(s y) ds = 0 = \underset{y\rightarrow +\infty}{\lim} \int_{\mathbb{R}}\!\! f(s) \sin(s y) ds
\end{equation}
for any $L^1$ integrable function $f(s)$ in $\mathbb{R}$. Additionally, as a corollary of
\begin{equation*}
  \int_{-\infty}^{+\infty}\!\! e^{\ima s y} dy = 2\pi \delta(s)
\end{equation*}
it holds that
\begin{equation} \label{deltafromsin}
  \underset{y\rightarrow +\infty}{\lim}\frac{\sin(s y)}{s} = \pi \delta(s)
\end{equation}
where $\delta(s)$ is the Dirac delta function and \eqref{deltafromsin} is to be understood as
\begin{equation*}
  \underset{y\rightarrow +\infty}{\lim} \int_{\mathbb{R}}\!\! f(s)\frac{\sin(s y)}{s} ds = \pi f(0),
\end{equation*}
$f(s)$ being an appropriate test function for the needs of standard distribution theory (infinitely differentiable with compact support). With the help of \eqref{RimLebL}, \eqref{deltafromsin} and a change of variable $y=\ln x$ we can thus write (always in a distributional sense \cite{Bielski})
\begin{equation} \label{limApos}
  \underset{y\rightarrow +\infty}{\lim} \mathcal{A}_{+}(y) = \pi \left[ \left(\alpha_{k,p}^{*}\alpha_{k,s}+\beta_{k,p}^{*}\beta_{k,s}\right) \delta(p-s) + \left(\alpha_{k,p}^{*}\beta_{k,s}+\beta_{k,p}^{*}\alpha_{k,s}\right) \delta(p+s)\right] .
\end{equation}

By turning to the case $x<0$ the corresponding expression for $\mathcal{A}(x)$ at the $x<<-1$ limit is written
\begin{equation*}
  \begin{split}
    \mathcal{A}(x)\big|_{x<<-1}=:\mathcal{A}_{-}(x)  =  & \ima (p+s) \cos [(p-s)\ln (-x)] \left(\eta_{k,p}^{*}\eta_{k,s} - \zeta_{k,p}^{*}\zeta_{k,s}\right) \\
    & + \ima (p-s) \cos[(p+s)\ln (-x)] \left(\eta_{k,p}^{*} \zeta_{k,s} - \zeta_{k,p}^{*}\eta_{k,s}\right) \\
    & - (p+s) \sin[(p-s)\ln (-x)] \left(\zeta_{k,p}^{*}\zeta_{k,s} + \eta_{k,p}^{*}\eta_{k,s}\right) \\
    & - (p-s) \sin[(p+s)\ln (-x)] \left(\zeta_{k,p}^{*} \eta_{k,s} + \eta_{k,p}^{*}\zeta_{k,s}\right) .
  \end{split}
\end{equation*}
and with a change of variable $y= \ln(-x)$ leads to
\begin{equation} \label{limAneg}
  \underset{y\rightarrow +\infty}{\lim} \mathcal{A}_{-}(y) =- \pi \left[ \left(\zeta_{k,p}^{*}\zeta_{k,s}+\eta_{k,p}^{*}\eta_{k,s}\right) \delta(p-s) + \left(\zeta_{k,p}^{*}\eta_{k,s}+\eta_{k,p}^{*}\zeta_{k,s}\right) \delta(p+s)\right] .
\end{equation}

Finally, by inserting \eqref{limApos} and \eqref{limAneg} in \eqref{genint1} we deduce that
\begin{equation*}
  \begin{split}
    \int_{-\infty}^{+\infty}\!\! P^k_{\ima s-1/2}(\ima\, x) (P^k_{\ima p-1/2}(\ima\, x))^{*} dx = & \pi \Big[\delta(p-s) (\alpha_{k,p}^{*}\alpha_{k,s}+\beta_{k,p}^{*}\beta_{k,s}+\zeta_{k,p}^{*}\zeta_{k,s} + \eta_{k,p}^{*} \eta_{k,s}) \\
    & +\delta(p+s) (\alpha_{k,p}^{*}\beta_{k,s}+\beta_{k,p}^{*}\alpha_{k,s}+\zeta_{k,p}^{*}\eta_{k,s} + \eta_{k,p}^{*} \zeta_{k,s}) \Big]
  \end{split}
\end{equation*}
and by substitution of \eqref{abks} and \eqref{zeks} we derive the following orthogonality relation
\begin{equation} \label{ApporthcontP}
  \int_{-\infty}^{+\infty}\!\!   P^k_{\ima s-1/2}(\ima\, x) (P^k_{\ima p-1/2}(\ima\, x))^{*}  dx = A(p,s)\delta(p-s)+ A(p,-s) \delta (p+s)
\end{equation}
where
\begin{equation} \label{coefAofP}
  \begin{split}
    A (p,s) = \cosh [(p+s)\frac{\pi}{2}] \Bigg[ & \frac{2^{-\ima (p-s)} \Gamma(-\ima p) \Gamma(\ima s)}{\Gamma(\frac{1}{2}-k-\ima p)\Gamma(\frac{1}{2}-k+\ima s)} \\
     & +\frac{2^{\ima (p-s)} \Gamma(\ima p) \Gamma(-\ima s)}{\Gamma(\frac{1}{2}-k+\ima p)\Gamma(\frac{1}{2}-k-\ima s)}\Bigg] .
  \end{split}
\end{equation}
Hence, we have a symmetric expression under the change $s\rightarrow -s$ (or $p\rightarrow - p$). We remind here that both $s$ and $-s$ correspond to the same eigenvalue $\ell(\ell+1) = -\frac{1}{4}-s^2$.

\subsection{Orthogonality for $Q^k_{\ima s- 1/2}(\ima\, x)$}

The associate Legendre function of the second kind, for $|z|>1$ can be written as \cite{Wolf}
\begin{equation} \label{asymptoticQ}
  \begin{split}
    Q^k_\ell (z) = \frac{2^{-(\ell+2)} e^{\ima k \pi} \sqrt{\pi} (1-z^2)^{k/2}}{\cos(\ell \pi) z^{k+\ell+1}} \Bigg[& \left(\cos[(k-\ell)\pi]+e^{\ima (\ell-k)\pi}\right) \frac{\Gamma(k+\ell+1)}{\Gamma(\ell+\frac{3}{2})} \\
    & + \ima\, 2^{2\ell+1}\sin[(k-\ell)\pi] \frac{\Gamma(k-\ell)}{\Gamma(\frac{1}{2}-\ell)} + O\left(\frac{1}{z}\right)\Bigg] .
  \end{split}
\end{equation}
Again we discriminate two cases after setting $z= \ima\, x$ and $\ell = \ima\, s- 1/2$:
\begin{itemize}
  \item  Case $x>>1$: By the approximation $1+\ima\, x \simeq \ima x$ and working in a similar manner to the previous section it is easy to derive
      \begin{equation}\label{Qpos}
        Q^k_{\ima s-1/2} (\ima\, x) \simeq \frac{1}{\sqrt{x}} \left( \tilde{\alpha}_{k,s} x^{\ima s} +\tilde{\beta}_{k,s} x^{-\ima s}  \right)
      \end{equation}
      with
      \begin{subequations}
        \begin{align}
          \tilde{\alpha}_{k,s} & = \frac{2^{\ima s}}{\sqrt{8\pi}} e^{-\frac{s\pi}{2}} e^{-\ima (k+\frac{1}{2})\frac{\pi}{2}} \coth(s\pi) \frac{\Gamma(k+\frac{1}{2}-\ima s)}{\Gamma(1-\ima s)} \\
          \tilde{\beta}_{k,s} & = \frac{2^{-\ima s}}{\sqrt{8\pi}}  e^{\frac{s\pi}{2}} e^{-\ima (k+\frac{1}{2})\frac{\pi}{2}} \left(2-\coth(s\pi)\right) \frac{\Gamma(k+\frac{1}{2}+\ima s)}{\Gamma(1+\ima s)}
        \end{align}
      \end{subequations}
  \item Case $x<<-1$: By an analogous process we are led to
     \begin{equation}\label{Qneg}
        Q^k_{\ima s-1/2} (\ima\, x) \simeq \frac{1}{\sqrt{x}} \left( \tilde{\zeta}_{k,s} (-x)^{\ima s} +\tilde{\eta}_{k,s} (-x)^{-\ima s}  \right)
      \end{equation}
      where
      \begin{subequations}
        \begin{align}
          \tilde{\zeta}_{k,s} & = \frac{2^{\ima s}}{\sqrt{8\pi}} e^{\frac{s\pi}{2}} e^{\ima (k+\frac{1}{2})\frac{\pi}{2}} \coth(s\pi) \frac{\Gamma(k+\frac{1}{2}-\ima s)}{\Gamma(1-\ima s)} \\
          \tilde{\eta}_{k,s} & = \frac{2^{-\ima s}}{\sqrt{8\pi}}  e^{-\frac{s\pi}{2}} e^{\ima (k+\frac{1}{2})\frac{\pi}{2}} \left(2-\coth(s\pi)\right) \frac{\Gamma(k+\frac{1}{2}+\ima s)}{\Gamma(1+\ima s)} .
        \end{align}
      \end{subequations}
\end{itemize}

Once more, for both cases we see that the solution in the form of $Q^k_{\ima s-1/2} (\ima\, x)$, satisfies the relevant boundary conditions as it decays with $x^{-1/2}$ when $x$ approaches infinity. Let us work now on the orthogonality relation by using \eqref{orthcond} in the form
\begin{equation} \label{genint2}
  \begin{split}
    \int_{-\infty}^{+\infty}\!\! Q^k_{\ima s-1/2}(\ima\, x) & (Q^k_{\ima p-1/2}(\ima\, x))^{*} dx = \\ & = \left[\frac{1+x^2}{p^2-s^2}\left((Q^k_{\ima p-1/2})^{*}\frac{d Q^k_{\ima s-1/2}}{dx}- Q^k_{\ima s-1/2} \frac{d (Q^k_{\ima p-1/2})^{*}}{dx}\right)\right]_{-\infty}^{+\infty} \\
    & = \Big[\frac{1}{p^2-s^2}\mathcal{B}(x)\Big]_{-\infty}^{+\infty} .
  \end{split}
\end{equation}
For large positive arguments we can express $\mathcal{B}(x)$ as
\begin{equation*}
  \begin{split}
    \mathcal{B}(x)\big|_{x>>1}=:\mathcal{B}_{+}(x)  = & \ima (p+s) \cos [(p-s)\ln x] \left(\tilde{\alpha}_{k,p}^{*}\tilde{\alpha}_{k,s} - \tilde{\beta}_{k,p}^{*}\tilde{\beta}_{k,s}\right) \\
    & + \ima (p-s) \cos[(p+s)\ln x] \left(\tilde{\alpha}_{k,p}^{*} \tilde{\beta}_{k,s} - \tilde{\beta}_{k,p}^{*} \tilde{\alpha}_{k,s}\right) \\
    & + (p+s) \sin[(p-s)\ln x] \left(\tilde{\alpha}_{k,p}^{*} \tilde{\alpha}_{k,s} + \tilde{\beta}_{k,p}^{*} \tilde{\beta}_{k,s}\right) \\
    & + (p-s) \sin[(p+s)\ln x] \left(\tilde{\alpha}_{k,p}^{*} \tilde{\beta}_{k,s} + \tilde{\beta}_{k,p}^{*} \tilde{\alpha}_{k,s}\right),
  \end{split}
\end{equation*}
while for large, in absolute value, negative $x$ we have
\begin{equation*}
  \begin{split}
    \mathcal{B}(x)\big|_{x<<-1}=:\mathcal{B}_{-}(x)  =  & \ima (p+s) \cos [(p-s)\ln (-x)] \left(\tilde{\eta}_{k,p}^{*}\tilde{\eta}_{k,s} - \tilde{\zeta}_{k,p}^{*} \tilde{\zeta}_{k,s}\right) \\
    & + \ima (p-s) \cos[(p+s)\ln (-x)] \left(\tilde{\eta}_{k,p}^{*} \tilde{\zeta}_{k,s} - \tilde{\zeta}_{k,p}^{*} \tilde{\eta}_{k,s}\right) \\
    & - (p+s) \sin[(p-s)\ln (-x)] \left(\tilde{\zeta}_{k,p}^{*}\tilde{\zeta}_{k,s} + \tilde{\eta}_{k,p}^{*} \tilde{\eta}_{k,s}\right) \\
    & - (p-s) \sin[(p+s)\ln (-x)] \left(\tilde{\zeta}_{k,p}^{*} \tilde{\eta}_{k,s} + \tilde{\eta}_{k,p}^{*} \tilde{\zeta}_{k,s}\right) .
  \end{split}
\end{equation*}
Once more the application of \eqref{RimLebL} and \eqref{deltafromsin} leads to
\begin{align} \label{limBpos}
  &\underset{y\rightarrow +\infty}{\lim} \mathcal{B}_{+}(y) = \pi \left[ \left(\tilde{\alpha}_{k,p}^{*}\tilde{\alpha}_{k,s}+ \tilde{\beta}_{k,p}^{*} \tilde{\beta}_{k,s}\right) \delta(p-s) + \left(\tilde{\alpha}_{k,p}^{*} \tilde{\beta}_{k,s}+ \tilde{\beta}_{k,p}^{*} \tilde{\alpha}_{k,s}\right) \delta(p+s)\right] \\ \label{limBneg}
  &\underset{y\rightarrow +\infty}{\lim} \mathcal{B}_{-}(y) =- \pi \left[ \left(\tilde{\zeta}_{k,p}^{*} \tilde{\zeta}_{k,s}+ \tilde{\eta}_{k,p}^{*}\tilde{\eta}_{k,s}\right) \delta(p-s) + \left(\tilde{\zeta}_{k,p}^{*} \tilde{\eta}_{k,s}+ \tilde{\eta}_{k,p}^{*} \tilde{\zeta}_{k,s}\right) \delta(p+s)\right]
\end{align}
where we have considered a new variable $y=\ln |x|$ in each case. As a result we have the following orthogonality relation
\begin{equation} \label{orthfinal2}
  \int_{-\infty}^{+\infty}\!\!   Q^k_{\ima s-1/2}(\ima\, x) (Q^k_{\ima p-1/2}(\ima\, x))^{*}  dx = B_1(p,s)\delta(p-s)+ B_2(p,-s) \delta (p+s)
\end{equation}
with
\begin{equation*}
  \begin{split}
    B_1(p,s) = & \frac{\pi}{4} \cosh[\frac{(p+s)\pi}{2}]  \Bigg[ 2^{-\ima (p-s)} \coth(p\pi) \coth(s\pi) \frac{\Gamma(k+\frac{1}{2}+\ima p) \Gamma(k+\frac{1}{2}- \ima s)}{\Gamma(1+\ima p) \Gamma(1-\ima s)}  \\
     & + 2^{\ima (p-s)} \left(2-\coth(p \pi)\right)\left(2-\coth(s\pi)\right)\frac{\Gamma(k+\frac{1}{2}-\ima p)\Gamma(k+\frac{1}{2}+\ima s)}{\Gamma(1-\ima p)\Gamma(1+\ima s)}\Bigg]
  \end{split}
\end{equation*}
and
\begin{equation*}
  \begin{split}
    B_2(p,s) = & \frac{\pi}{4} \cosh[\frac{(p-s)\pi}{2}] \Bigg[ 2^{-\ima (p+s)} \coth(p\pi) \left(2-\coth(s\pi)\right) \frac{\Gamma(k+\frac{1}{2}+\ima p) \Gamma(k+\frac{1}{2}+\ima s)}{\Gamma(1+\ima p) \Gamma(1+\ima s)} \\
    & + 2^{\ima (p+s)} \coth(s\pi) \left(2-\coth(p\pi)\right) \frac{\Gamma(k+\frac{1}{2}-\ima p) \Gamma(k+\frac{1}{2}-\ima s)}{\Gamma(1-\ima p)\Gamma(1-\ima s)} \Bigg].
  \end{split}
\end{equation*}
As we can see \eqref{orthfinal2} possesses no parity symmetry $p\rightarrow -p$ or $s\rightarrow -s$. Henceforth, even though $s$ and $-s$, for example, lead to the same eigenvalue $\ell(\ell+1)=-\frac{1}{4}-s^2$, they have different probability amplitudes in contrast to what happens if one admits the $P^k_{\ima s-1/2}(\ima x)$ solution.

\begin{acknowledgments}

N. D. acknowledges financial support by FONDECYT postdoctoral grant no. 3150016.

\end{acknowledgments}

\end{document}